\documentclass{iopart}
\usepackage{graphicx}

\newcommand{\be}{\begin{equation}}
\newcommand{\en}{\end{equation}}
\begin{document}

\title[Anderson localization transition in ensemble of ultrametric random matrices]{Anderson localization transition and eigenfunction multifractality in ensemble of ultrametric random matrices.}

\author{Yan V Fyodorov$^1$, Alexander Ossipov$^1$, Alberto Rodriguez$^2$}
\address{$^1$ School of Mathematical Sciences, University of Nottingham, Nottingham NG72RD, England}
\address{$^2$ Department of Physics and Centre for Scientific Computing, University of Warwick, Coventry CV47AL, United Kingdom}

\date{\today}

\begin{abstract}
 We demonstrate that by considering disordered single-particle Hamiltonians (or their random matrix versions)
on ultrametric spaces one can generate an interesting class of models exhibiting Anderson metal-insulator transition.
We use the weak disorder virial expansion to determine the critical value of the parameters and to calculate the values of the multifractal exponents for inverse participation ratios. Direct numerical simulations agree favourably with the analytical predictions.
\end{abstract}

\pacs{72.15.Rn, 71.30.+h, 05.45.Df}


\maketitle

A metric space is called {\it ultrametric}, if a distance function $d(x,y)$ satisfies the strong triangle inequality $d(x,y)\le \max\{d(x,z), d(z,y)\}$ for any three points $x$, $y$ and $z$, so that all triangles are either isosceles or equilateral.  The notion of ultrametricity appears naturally in $p$-adic analysis \cite{p-adic} which plays important role in various branches of mathematics and string theory. Ultrametric spaces have found many useful applications in physics and beyond \cite{RTV86},  most notably in the theory of spin glasses \cite{sglass}
where they describe the hierarchical organization of free energy landscapes.
Random matrices whose structure reflects some specific aspects of ultrametricity were under active study recently
due to their relevance to description of relaxation properties in complex landscapes typical for heteropolymers and related systems \cite{Nechaev}.

The aim of the present work is to demonstrate that by considering disordered single-particle Hamiltonians (or their random matrix versions)
on ultrametric spaces one can generate an interesting class of models exhibiting Anderson metal-insulator transition. The latter is one of the central phenomena in the theory of disordered quantum systems and describes an abrupt change in the nature of eigenstates from extended to spatially localized. Exactly at the transition point eigenstates are very inhomogeneous and strongly fluctuating  and can be characterized by non-trivial multifractal dimensions. Despite a great progress in understanding of the Anderson transition achieved in the last
 decades \cite{EM08} based mainly on the mapping to the nonlinear $\sigma-$model \cite{Wegner,Efetov}, there are still relatively few models of disordered Hamiltonians exhibiting the
 localisation transition and affordable for direct analytical studies. Among those deserving special mentioning are the models on the infinite tree-like graph (Bethe lattice) \cite{Efetov,BL} and the ensemble of power-law banded random matrices introduced in \cite{PRBM} and further investigated in \cite{EM00} which became rather popular in recent years. The problem of rigorous mathematical analysis of the Anderson transition is yet outstanding, see e.g. \cite{DSZ}.

 The idea that models of statistical mechanics on hierarchical spaces with effectively ultrametric structure can be efficiently
  studied analytically due to their high symmetry goes back to Dyson\cite{Dyson} and proved to be rather useful in shedding light
 on renormalisation group methods and developing rigorous mathematical proofs in general theory of phase transitions \cite{CE}.
 Recently there was a systematic attempt to put disordered models of spin-glass type on such lattices \cite{SG}.
 We hope that the models proposed in the present paper will provide a useful framework for further insights
 into the nature of the Anderson localisation transition, and perhaps could yield to a rigorous mathematical analysis.

Let us start by introducing a random matrix ensemble of Hermitian matrices with ultrametric structure
whose eigenvectors exhibit the Anderson transition when changing some control parameter. The idea behind such an approach is that ensembles of that sort can be not only efficiently studied numerically, but 
 should be also amenable to a variety of existing analytical techniques like, for example, mapping to a kind of nonlinear $\sigma-$model on ultrametric space, which could keep promise for an exact solution (see a short discussion in the end  of the paper). 
 Leaving the investigation of these issues to a future research, in this paper we are to exploit some recent analytical insights provided by works on almost-diagonal random matrices \cite{YK03,YO07}.

\begin{figure}[tb]
(a) \includegraphics[width=0.4\columnwidth]{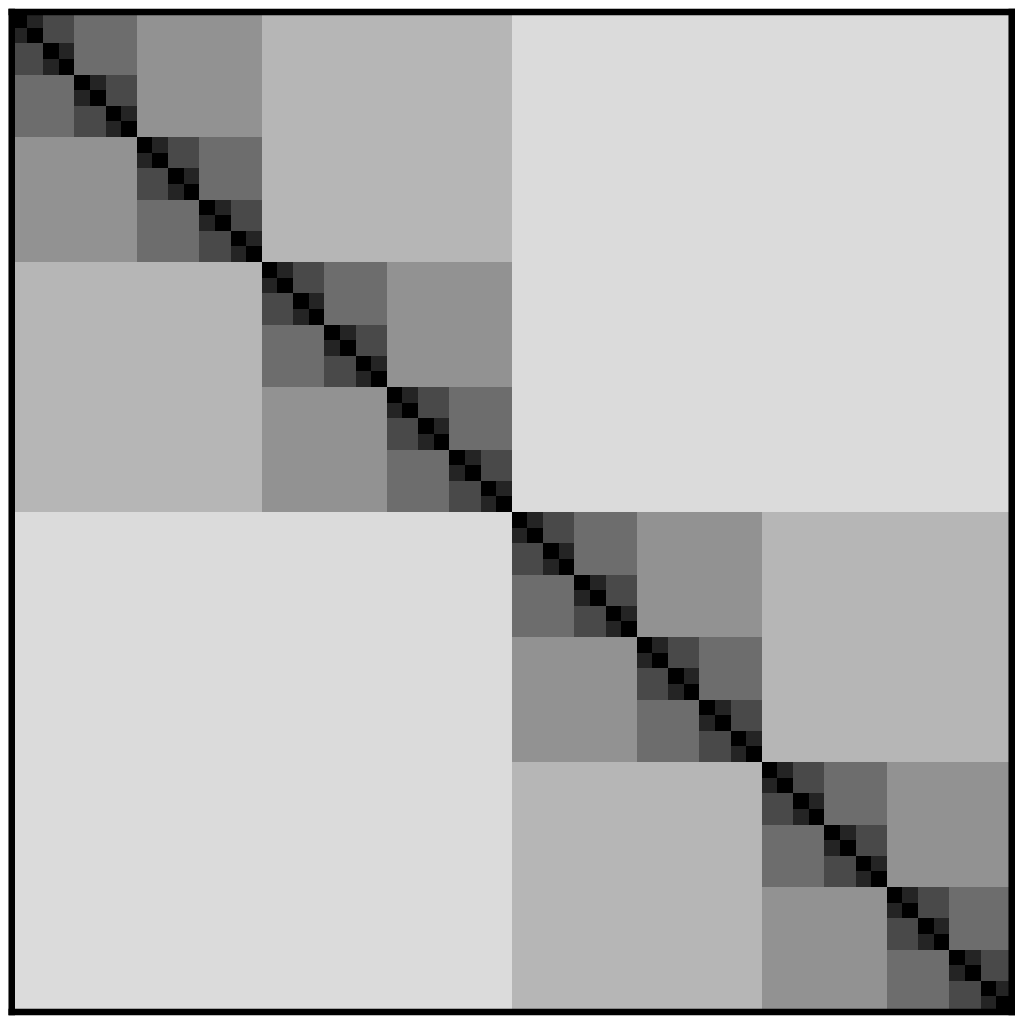}
(b) \includegraphics[scale=0.75]{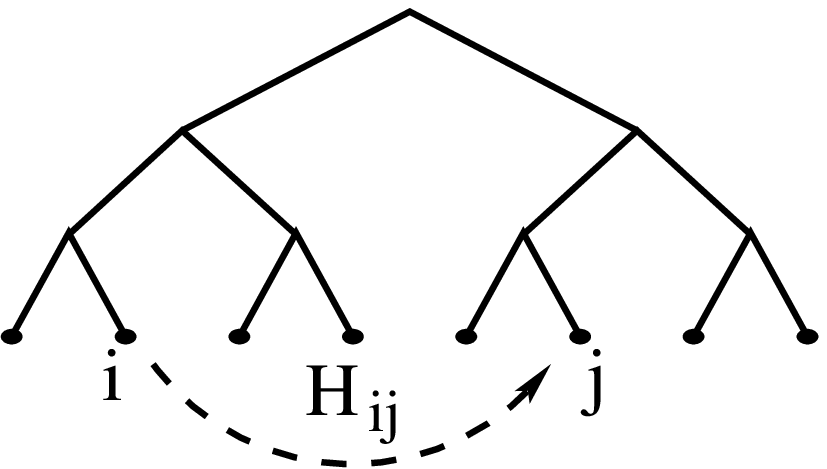}
\caption{\label{fig_structure} (a) Schematic structure of variances of entries of an ultrametric random matrix. (b) Hopping between nodes of the ultrametric lattice.}
\end{figure}

The ultrametric structure for a particular representative $H$ of the suggested ensemble of random matrices  is encoded in the choice of variances for the Gaussian-distributed entries. Namely, all $2^K\times 2^K$ entries $H_{ij}$ of the matrix $H$ are taken as independent complex Gaussian random variables (up to the obvious Hermiticity constraint $H_{ij}^{\ast}=H_{ji}$) with the zero mean value. All diagonal elements of $H$ are real and have the same variance: $\left\langle H_{ii}^2\right\rangle=W^2,\, \forall i$. The variances of the off-diagonal elements have a block structure typical for "hierarchic" lattices (Fig.~\ref{fig_structure}a), with blocks of the sizes $1,2,2^2,\ldots,2^{K-1}$. Inside each block of the size $2^l$ the variances $\left\langle H_{ij}^{\ast}H_{ij}\right\rangle $ are constant and chosen to be $J^2/p^{2l}$, controlled by two real parameters $J$ and $p$. Such a random matrix describes random hopping between {\it boundary nodes} of a tree of $K$ generations with given coordination number which we choose for simplicity to be equal to two (Fig.~\ref{fig_structure}b) . The underlying tree serves the purpose of defining the ultrametric distance  $d(i,j)$ between any pair of boundary nodes. It is defined as the number of edges in the shortest path connecting nodes $i$ and $j$. As easy to check the corresponding distance is ultrametric. The formula for the variances can be then expressed naturally in terms of $d(i,j)$ as $\left\langle H_{ij}^{\ast}H_{ij}\right\rangle=J^2/p^{d(i,j)-2}$.

The ultrametric random matrix model introduced above exhibits the Anderson transition by changing parameter $p$. Namely,
 in the thermodynamic limit  $K\to \infty$ the eigenvectors of $H$ are extended for $p<2$, localized for $p>2$ and critical (multifractal) at $p=2$ . We expect this scenario to hold for any ratio $J/W$, implying in particular that at $p=2$ changing $J/W$ should allow to pass from strongly to weakly multifractal eigenstates. Assuming that $J/W\ll 1$,  the fractal dimensions $d_q$ of the eigenvectors (defined below in Eqs.(\ref{moments_def}-\ref{scaling})) in the critical regime can be calculated explicitly and are given by:
\be\label{fractal_dim}
d_q=\frac{J}{W}\frac{\sqrt{\pi}}{\sqrt{2}\ln 2}\frac{\Gamma (q-1/2)}{\Gamma (q)},\quad q>\frac{1}{2},
\en
where $\Gamma (q)$ is the Euler Gamma-function.

We define the moments of the eigenvector components $I_q$ (also known as inverse participation ratios, IPR's) in the standard way:
\begin{eqnarray}\label{moments_def}
I_q(E)&=&\frac{1}{N}\sum_{n=1}^N P_q(E,n) \nonumber\\
P_q(E,n)&=&\frac{1}{\rho(E)}\sum_{k=1}^N|\Psi_k(n)|^{2q}\delta (E-E_k),
\end{eqnarray}
where  the normalized eigenvectors $\Psi_k$ with $N\equiv 2^K$ components $\Psi_k(n),\,n=1,2,\ldots, N$ correspond to the eigenvalues $E_k$
and $\rho (E)$ stands for the mean density of states at the energy $E$ (which will be fixed to zero for simplicity).
At criticality the ensemble-averaged moments of the eigenvectors scale anomalously with the system size $N$, defining a set of fractal dimensions $d_q$:
\be\label{scaling}
\left\langle I_q\right\rangle\propto N^{-d_q(q-1)}.
\en

In order to calculate the moments of the eigenvectors under the condition $J/W\ll 1$ we use the method of the virial expansion developed in \cite{YO07, KOY09}. The methods allows us to construct a well-controlled perturbation expansion of the moments of the eigenvectors using the small parameter  $J/W$. The first term of the expansion is generated by the diagonal part of a matrix $H$. All corresponding eigenvectors have only one non-zero component, which is equal to one due to the normalization condition. As a result all moments of the eigenvectors are equal to one in this approximation: $\left\langle I_q^{(0)} \right\rangle=1$. The corrections to this result originating from the off-diagonal matrix elements of $H$ can be represented as certain integrals over commuting and anti-commuting variables. That integrals can be either calculated explicitly \cite{YO07, KOY09} or alternatively they can be interpreted as the moments of the eigenvectors of certain submatrices of a matrix $H$. To be specific, the lowest order in $J/W$ correction $I_q^{(1)}$ is generated by the  following $2\times 2$ sub-matrices of $H$:
\begin{eqnarray}\label{E_2}
H(n,m) & = & \left(\begin{array}{cc}
H_{nn} & H_{nm} \\ H_{nm}^{\ast} & H_{mm}
\end{array} \right),\nonumber  n,m=1,\dots,N,\quad n\neq m
\end{eqnarray}

Denoting by $P_q(n,m)$ the moments of the first component of the corresponding eigenvectors we are able to represent $\left\langle I_q^{(1)}\right\rangle$ as follows:
\be\label{mom_two}
\left\langle I_q^{(1)}\right\rangle=\frac{1}{N}\sum_{n\neq m}^N  (\left\langle P_q(n,m)\right\rangle -1)
\en
The subtraction of unity is required in order to get rid of the contribution already taken into account at the previous step. The above approximation implies that the pair-wise "interaction" of the levels of the unperturbed system is taken into account at this stage. Essentially the same idea was underlying the real-space renormalization approach developed for critical systems with a long-range interaction  and the power-law random banded matrix model \cite{Lev90,EM00}.

Below we calculate the averaged value of $I_q^{(1)}$ using an explicit expression for the eigenvectors components. To this end let us introduce the following notation $E_1\equiv H_{nn}$, $E_2\equiv H_{mm}$, $h\equiv H_{nm}$ and $P_q\equiv P_q(n,m)$. The eigenvalues $\lambda_1$, $\lambda_2$ of the matrix $H(n,m)$ are equal to
\be\label{lambda}
\lambda_{1,2}=\frac{1}{2}(E_1+E_2\pm\sqrt{(E_1-E_2)^2+4|h|^2}).
\en
The absolute values of the components $\psi_{1,2}$ of the corresponding normalised eigenvectors are given by
$|\psi_{1,2}|= \frac{|h|}{\sqrt{|h|^2+(\lambda-E_{1,2})^2}}$. Averaging the moments of $|\psi_{1}|$ first over  $E_1$ and $E_2$ we obtain the following expression for $\left\langle P_q\right\rangle_E$:
\be\label{IPR-diff}
\left\langle P_q\right\rangle_E=\int\int_{-\infty}^{\infty}
\frac{dE_1dE_2}{2\pi\rho W^2}\sum_{i=1}^2
        \frac{e^{-\frac{E_1^2+E_2^2}{2W^2}}|h|^{2q}\delta (\lambda_i)}{(|h|^2+(\lambda_i-E_1)^2)^q},
\en
where $\rho$ is the density of states at $E=0$. The latter can be easily calculated in the diagonal approximation giving the result $\rho=1/\sqrt{2\pi}W$. The corrections generated by off-diagonal elements of $H$ is of the order of $(J/W)^2$ \cite{YK03} and hence can be neglected. Eq.(\ref{lambda}) shows that $\lambda_1=0$ if $E_1E_2=|h|^2$ and $E_1+E_2\le 0$, while $\lambda_2=0$ if $E_1E_2=|h|^2$ and $E_1+E_2\ge 0$. Therefore we should consider separately two domains of integration $E_1+E_2\le 0$ and $E_1+E_2\ge 0$. In each domain only one of the $\delta$-functions in Eq.(\ref{IPR-diff}) survives.
After a simple change of variables we are able to integrate over $E_1$ in Eq.(\ref{IPR-diff}):
\begin{eqnarray}
\left\langle P_q\right\rangle_E
      &=&\int_{-\infty}^{\infty}  \frac{dE_2}{\sqrt{2\pi}W}e^{-\frac{|h|^4+E_2^4}{2W^2E_2^2}}
      \left(\frac{E_2^2}{E_2^2+|h|^2}\right)^{q-1}=\nonumber\\
&=&\int_{-\infty}^{\infty} \frac{dx\:|h|}{\sqrt{2\pi}W} e^{-\frac{|h|^2}{2W^2}\left(\frac{1}{x^2}+x^2\right)}
      \left(\frac{x^2}{x^2+1}\right)^{q-1},
\end{eqnarray}
where the new variable of integration $x=|h|E_2$ is introduced. Now we are ready to perform averaging over the complex variable $h$. Since the expression above depends only on $|h|$ the integration over the argument $h$ is trivial and just leads to appearance of the factor $2\pi$ while integration over $|h|$ can be done explicitly:
\be\label{int}
\left\langle P_q\right\rangle_{E,h}
       =\frac{\alpha}{2}\int_{-\infty}^{\infty} dx
      \frac{\left(x^{-2}+1\right)^{1-q}}
      {\left(1+\alpha\left(x^{-2}+x^2\right)\right)^{3/2}},
\en
where $\alpha\equiv \sqrt{\left\langle |h|^2\right\rangle}/\sqrt{2}W$. The asymptotic behaviour of this expression in the limit $\alpha\to 0$ has the following form:
\be\label{expan}
\left\langle P_q\right\rangle_{E,h}=1-\sqrt{\pi}\frac{\Gamma (q-1/2)}{\Gamma (q-1)}\alpha+O(\alpha^2).
\en
The first term in the expansion corresponds to the diagonal approximation and is cancelled by $-1$ in Eq.(\ref{mom_two}), while the second one gives a non-trivial contribution. Collecting the results of Eqs.(\ref{moments_def},\ref{mom_two},\ref{expan}) we find $\left\langle I_q\right\rangle$ up to the first order in $J/W$:
\begin{eqnarray}\label{mom_general}
\left\langle I_q\right\rangle&=&1-\sqrt{\frac{\pi}{2}}\frac{\Gamma (q-1/2)}{\Gamma (q-1)}\frac{1}{N}\sum_{n\neq m}^N\frac{\sqrt{\left\langle|H_{nm}|^2\right\rangle}}{W}\\
&=& \label{mom_ultra}1-\frac{J}{W}\sqrt{\frac{\pi}{2}}\frac{\Gamma (q-1/2)}{\Gamma (q-1)}
\sum_{m=2}^N\frac{1}{p^{d(1,m)/2-1}}.
\end{eqnarray}
According to the definition of the ultrametric distance $d(n,m)$ the sum in Eq.(\ref{mom_ultra}) is given by the geometric series:
\be\label{sum}
S\equiv \sum_{m=2}^N\frac{1}{p^{d(1,m)/2-1}}=\sum_{i=0}^{K-1}\left(\frac{2}{p}\right)^i
=\frac{\left(\frac{2}{p}\right)^K-1}{\frac{2}{p}-1}.
\en
As $K\to \infty$ the asymptotic behaviour of the above sum crucially depends on the value of the  parameter $p$:
 for $p>2$ the sum tends to the finite value $S\to p/(p-2)<\infty$ independent of $N$  implying that eigenvectors are localized, whereas
in opposite situation $p<2$ we see that $S$ diverges as $S\approx (p/(2-p)) N^{1-\log_2{p}}$, signalling of a breakdown of the perturbation
theory for arbitrarily weak off-diagonal coupling. It is natural to interpret such breakdown as the signal of eigenvector delocalization.
 Finally, at the critical point
$p=2$ we have $S=K=\log_2N$ leading to
\be
\left\langle I_q\right\rangle= 1-\frac{J}{W}\sqrt{\frac{\pi}{2}}\frac{\Gamma (q-1/2)}{\Gamma (q-1)}\log_2N.
\en
Comparing this result with the definition of $d_q$ given by Eq.(\ref{scaling}), and expecting $d_q\ll 1$ for $J/W\ll 1$
we interpret this marginal behaviour  in terms of the anomalous scaling of the
moments of the eigenvectors, and the expression for the fractal dimensions (\ref{fractal_dim}) immediately follows.
The condition $q>1/2$ is necessary to guarantee the convergence of the derivative with respect to $\alpha$ of the integral in (\ref{int}) at $\alpha=0$.
One expects that the region $q<1/2$ can be covered by employing the powerful symmetry relation discovered in \cite{symm} , see below.

One can notice that the $q$-dependence of the above expression is exactly the same as the one found for the ensemble of the power-law random banded matrices \cite{EM00}. The matrix elements of that ensemble are independently distributed Gaussian variables with zero mean and with variance decaying in a power-law fashion as a function of the distance to the main diagonal. Such a coincidence can be easily explained by Eq.(\ref{mom_general}). Indeed, we can see that $q$-dependence enters only in the pre-factor and thus is always the same for {\it any} Gaussian ensemble
with independent entries. However we would like to stress that this "superuniversal" behaviour should be expected only in the regime $J/W\ll 1$.

\begin{figure}[tb]
\includegraphics[width=\columnwidth]{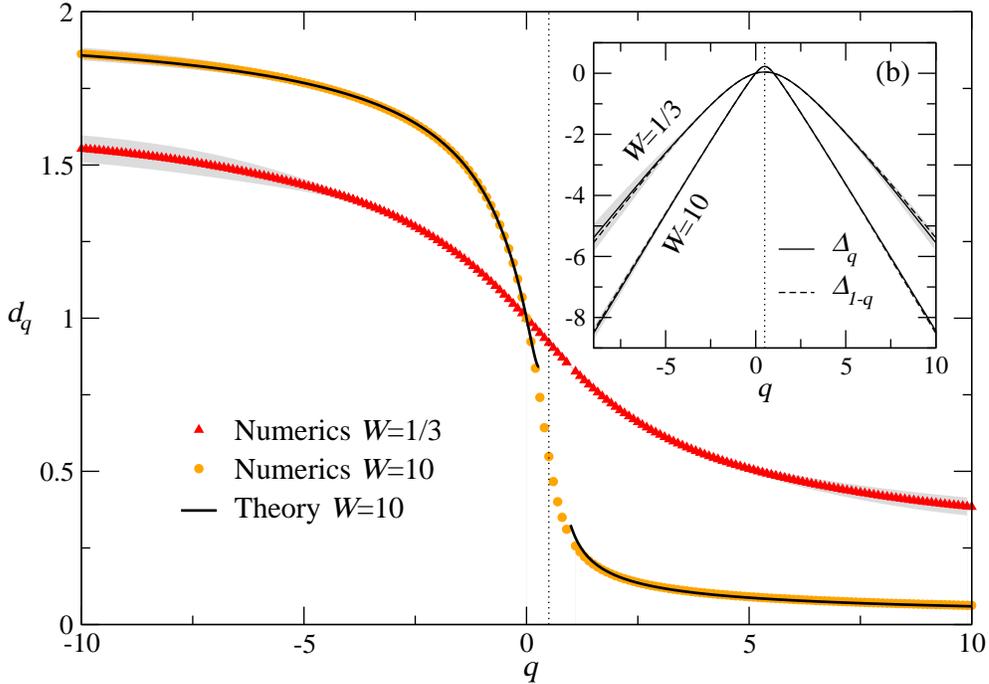}
\caption{\label{fig_dq} Fractal dimensions $d_q$ for $J=1$, $W=10$ (circles) and $W=1/3$ (triangles). The solid line is the analytical prediction based on Eq.(\ref{fractal_dim}). Inset: numerically calculated exponents $\Delta_q$ (solid) and $\Delta_{1-q}$ (dashed). The gray shaded background marks the 95\% confident interval of the numerical results.}
\end{figure}

The results of numerical simulations confirming our predictions are presented in Fig.~\ref{fig_dq}. The moments of the eigenvectors, whose eigenvalues are close to $E=0$, were calculated for various system sizes ranging from $N=2^{8}$ to $N=2^{14}$. To obtain their averaged values reliably a large number of random matrices were generated, so that the number of eigenvectors contributing to the computed averaged values were ranged from $10^3$ for $N=2^{14}$ up to $10^5$ for $N=2^{8}$. The extracted values  of the  fractal dimensions $d_q$ for $J=1$, $W=10$ and $W=1/3$ are represented in Fig.~\ref{fig_dq} by circles and triangles respectively. For $q<0$ the eigenstates intensities were first coarse-grained over boxes containing $8$ sites. The values of $d_q$ for $W=10$ are in perfect agreement with our analytical prediction (solid line in Fig.~\ref{fig_dq}). The latter was calculated directly from Eq.(\ref{fractal_dim}) for $q>1/2$. To make the prediction for $q<1/2$ we used the remarkable symmetry relation discovered in \cite{symm} and intimately related to profound fluctuation relations in the field of non-equilibrium dynamics \cite{MBC}. According to it the anomalous multifractal exponent defined as $\Delta_q\equiv(d_q-d)(q-1)$, where $d$ is the dimensionality of the space, is symmetric with respect to line $q=1/2$, i.e. $\Delta_q=\Delta_{1-q}$. In our model $d=1$, as according to Eq.(\ref{scaling}) the fractal dimensions are defined by the scaling of the moments of the eigenvectors with respect to the total number of sites $N$ (``volume'' of the system). We checked the validity of the relation for our model (inset of Fig.~\ref{fig_dq}) and employed it to calculate $d_q$ for $q<1/2$ analytically using Eq.(\ref{fractal_dim}).  The fact that the expression (\ref{fractal_dim}) diverges at $q=1/2$ indicates that our perturbative approach breaks down at some values $q_{\ast}$ close to $q=1/2$, at which Eq.(\ref{fractal_dim}) becomes of the order of one. However we would like to mention that the values $q_{\ast}$ can be made arbitrary close to $1/2$ by taking $J/W$ sufficiently small.

Another possible realization of the ultrametricity in the framework of disordered systems is the ultrametric Anderson model, which is a version of the Anderson model defined on the lattice of $N=2^K$ sites supplied with the ultrametric distance $d(i,j)$ described above. The diagonal matrix elements of the Hamiltonian are assumed to be random and characterized by the variance $W$. The off-diagonal matrix elements $H_{ij}$ of the Anderson model are however non-random. Their values $H_{ij}=J/p^{d(i,j)/2-1}$ depend on two parameters $J$ and $p$ and the distance $d(i,j)$ similar to corresponding variances of an ultrametric random matrix.

Although one can not apply directly our results to the Anderson model, one can use the argument due to Levitov \cite{Lev90} to establish the existence of the Anderson transition for this model. According to that argument,
two sites $i$ and $j$ will be mixed by the off-diagonal entries of ${\cal H}$ only when they are "in resonance": $|H_{ii}-H_{jj}|<H_{ij}$. It is easy to estimate the probability that two sites are in resonance:
\begin{equation}\label{3}
P_{ij}=\int\int\frac{dH_{ii}dH_{jj}}{2\pi W^2}e^{-H^2_{ii}/2W^2-H^2_{jj}/2W^2}
\theta\left(H_{ij}-|H_{ii}-H_{jj}|\right)
\end{equation}
where $\theta(x)$ is the step-function.  Calculating the integral  yields
$ P_{ij}=\sqrt{\frac{2}{\pi}}\int_{0}^{H_{ij}/\sqrt{2}W}e^{-x^2/2}\,dx\approx \sqrt{\frac{2}{\pi}}\frac{H_{ij}}{W}$
in our approximation. Then the mean total number of sites $\left\langle\# \right\rangle_i$ which are in resonance with a given site $i$ is given by:
\begin{equation}\label{4}
\left\langle\# \right\rangle_i=\sum_{j\ne i}P_{ij}=\sqrt{\frac{2}{\pi}}\frac{1}{W}\sum_{j\ne i}\,H_{ij}
\end{equation}
When  the number $\left\langle\# \right\rangle_i$ is divergent in the limit $N\to \infty$ as a power of $N$ the couplings, whatever weak, produce delocalization of eigenvectors in the site basis. On the other hand that number tending to a finite limit indicates localization of the corresponding eigenvectors in the weak coupling limit (which of course does not preclude possibility of delocalization at some critical coupling).
Finally, the transition between the two regimes happens precisely when $\left\langle\# \right\rangle_i\sim \ln{N}$.
For the ultrametric Anderson model $\sum_{j\ne i}H_{ij}/W=(J/W)S$, where $S$ is defined by Eq.(\ref{sum}). Thus we conclude that depending on
$p$ the nature of the eigenvectors of the Anderson model is the same as for its random matrix analogue: localized for $p>2$, extended for $p<2$ and critical at $p=2$, although we expect the dependence of anomalous dimensions $d_q$ on the parameter $J/W$ to be different in the two models.
The statistics of energy level spacings is expected to be Poisson for $p>2$ and Wigner-Dyson for $p<2$.
For earlier studies of spectral properties of the random Schroedinger-like operators with hierarchical structure see
 \cite{molchanov,Kri07}.

Finally, let us say a few words about possibilities to understand the opposite limit of strong off-diagonal coupling $J\gg W$. In the original formulation of the model neglecting diagonal entries does not seem to lead to any essential simplification, and we are unfortunately unable
to provide any insights to the problem. However, one can think of introducing an "n-orbital" variant of our model in the spirit of \cite{Wegner}.
 In this way one should replace each entry $H_{ij}$ of the hierarchical random matrix Hamiltonian 
 with $n\times n$ block of independent, identically distributed zero mean Gaussian elements
 with the variances $J^2/p^{2l}$ inherited from the original $n=1$  model. Letting $n\to \infty$ and simultaneously 
 rescaling the parameters $J$ and $W$ in such a way that $J^2/W^2 =t/n$ with some $t=O(1)$ ensures a non-trivial limit of the theory
 equivalent to a lattice version of the standard nonlinear $\sigma-$model, with couplings depending on the ultrametric distance between 
 the lattice sites.  In such approach the strong coupling limit $t\gg 1$ will be amenable to the standard perturbative expansion in diffusion-like 
modes like in \cite{PRBM}, with the diffusion propagator replaced by its analogue on the hierarchical lattice (note that as the model lacks the translation invariance the standard Fourier modes expansion is ineffective and should be replaced by a certain ultrametric analogue).
We leave analysis of this and related questions to subsequent publications.

AO acknowledges support from the EPSRC [EP/G055769/1]. YF is grateful to S.~Nechaev for
triggering his interest in hierarchical models. AR gratefully acknowledge funding from EPSRC (EP/C007042/1).

\end{document}